\documentclass[12pt]{iopart}
\usepackage{graphicx}
\usepackage{hyperref}
\hypersetup{pdfborder=0 0 0,colorlinks=true,citecolor=blue,linkcolor=blue}
\usepackage{csquotes}
\usepackage{graphicx}
\usepackage{csquotes}
\usepackage{wrapfig}
\hypersetup{pdfborder=0 0 0,colorlinks=true,citecolor=blue,linkcolor=blue}
\newcommand{\MX}{MX$_2$}
\newcommand{\MS}{MoS$_2$}
\newcommand{\TiS}{TiS$_2$}
\newcommand{\MSe}{MoSe$_2$}
\newcommand{\MTe}{MoTe$_2$}

\newcommand{\HfS}{HfS$_2$}
\newcommand{\HfSe}{HfSe$_2$}

\usepackage{iopams}  
\begin{document}
\title[1D metallic states at lateral heterojunctions]{1D metallic states at 2D transition metal dichalcogenide semiconductor heterojunctions}

\author{Sridevi Krishnamurthi, Geert Brocks}

\address{IComputational Materials Science, Faculty of Science and Technology and MESA+ Institute for Nanotechnology, University of Twente, the Netherlands. }
\ead{g.h.l.a.brocks@utwente.nl}
\vspace{10pt}
\begin{indented}
\item[]August 2020
\end{indented}

\begin{abstract}
Two-dimensional (2D) lateral heterojunctions of transition metal dichalcogenides (TMDCs) have become a reality in recent years. Semiconducting TMDC layers in their common $H$-structure have a nonzero in-plane electric polarization, which is a topological invariant. We show by means of first-principles calculations that lateral 2D heterojunctions between TMDCs with a different polarization  generate one-dimensional (1D) metallic states at the junction, even in cases where the different materials are joined epitaxially. The metallicity does not depend upon structural details, and is explained from the change in topological invariant at the junction. Nevertheless, these 1D metals are susceptible to 1D instabilities, such as charge- and spin-density waves, making 2D TMDC heterojunctions ideal systems for studying 1D physics.
\end{abstract}
\section*{Introduction}
Lateral structures made from two-dimensional (2D) materials have been gaining attention both in the experimental and theoretical domains, because of their promise to open the route towards truly 2D electronics. In-plane $p$-$n$ junctions and barrierless Schottky contacts between 2D compounds provide the basic building blocks of 2D electronic devices \cite{nmat4091,_valos_Ovando_2019,Zhang2018}. Lateral heterojunctions between various 2D semiconductors have been realized since chemical vapour deposition techniques have enabled the growth of sharp one-dimensional (1D) interfaces between different 2D materials. 

Transition metal dichalcogenides (TMDCs) are a particular versatile class of compounds in this respect, where 2D heterostructures of TMDCs with a similar crystal structure are grown routinely \cite{Huang2014,Levendorf,Li524}. In addition, 2D junctions between TMDCs with different crystal structures can be produced \cite{Lin2014}, and even structures with a large lattice constant mismatch, such as graphene or h-BN and {\MS}, have been grown as lateral junctions \cite{doi:10.1002/adma.201505070}.

Two-dimensional TMDCs, {\MX}, M transition metal, X = S, Se, Te, constitute an extensive family of compounds, covering metals and semiconductors, depending on their elemental composition, and their crystal phase. The compounds with M = Mo, W, and X = S, Se, have been at the center of attention, as they are direct band-gap semiconductors with potential applications in optoelectronics, photovoltaics, and photocatalysis \cite{doi:10.1002/jctb.6335,Jaramillo100,Ponraj_2016}. Not surprisingly then, the focus has been on junctions made from these semiconducting materials, their band alignments and interface transport properties  \cite{doi:10.1002/adma.201505070,Huang2014,Levendorf,Li524,Lin2014,nmat4091,_valos_Ovando_2019,Zhang2018}. 

What has been exploited much less is the notion that, in their most common structure, these semiconductors are materials with a net in-plane polarization. Moreover, insulating or semiconducting TMDCs  with transition metals from different elemental groups can have a different polarization. If one creates an in-plane heterostructure between two TMDCs with different polarization, then polarization charges develop along the junction, even if the structure at the junction is perfectly epitaxial. In a sufficiently large structure, these polarization charges have to be compensated by electrons or holes to avoid a polar catastrophy (a diverging potential) \cite{oxide}, which we will show in this paper results in 1D metallic states localized at the junction.  

TMDC grain boundaries and edges are known to have 1D metallic states that are exclusively localized at the boundary of edge, which display electronic properties that are especially prominent in 1D systems, such as charge density waves (CDWs), spin density waves (SDWs), or L\"{u}ttinger liquid behavior \cite{PhysRevLett.87.196803,doi:10.1021/acs.nanolett.5b02834,komsa-MTB,PhysRevB.87.205423,krishnamurthi2020}. As grain boundaries and edges constitute 1D extended defects, one might argue that the appearance of 1D states is necessarily connected to the rather drastic character of these defects.  

However, the presence or absence of 1D states is dictated by the bulk polarization of the corresponding TMDCs. In this paper we will show that they also appear in perfectly lattice-matched heterojunctions, and these states arise out of a difference in the bulk polarisation at the junction. We will discuss examples from two different cases, a polar/polar junction and a polar/non-polar junction. We also investigate possible perturbations that could lead to a band gap in these 1D metallic structures, such as charge ordering and/or spin ordering.

There have been computational studies on interfaces of semiconducting polar 2D materials such as AlN/SiC and ZnO/SiC \cite{PhysRevB.88.161411,PhysRevB.95.045302}, where these junctions have been found to be metallic. The junction between blue and black phosphorene has been predicted to have a charge density wave arising at the 1D metallic interface \cite{doi:10.1002/smll.201803040}. Clean junctions between such materials may be difficult to realize experimentally. Considering recent developments in growth techniques, junctions between TMDCs are more accessible.

\section*{Methods}
To model a 2D TDMC heterojunction, we build a supercell that has a width of twelve {\MX} unit cells, six for each of the two compounds forming the junction. This unit is periodically repeated in plane, such that each supercell contains two junctions, see Fig. \hyperref[figure1]{1}(a), for instance. Perpendicular to this 2D plane we use a vacuum spacing of 15 \AA\ to prevent an interaction between the periodic images.

We perform density functional theory (DFT) calculations, with the generalized gradient approximation (GGA) PBE and PBE+U functionals, and the projector augmented wave (PAW) method, using the Vienna ab initio simulation package (VASP) \cite{PhysRevB.54.11169,PhysRevB.59.1758,PhysRevB.50.17953,PhysRev.136.B864,PhysRev.140.A1133,PhysRevB.23.5048,PhysRevB.57.1505}. For the transition metals in {\MX} the outer $s$, $p$, and $d$ shells are treated as valence electrons, and for the chalcogen atoms the outer $s$ and $p$ shells. A cut-off of 400 eV for the kinetic energy of the plane waves, and a $k$-point sampling of 12 points per unit cell along the direction of the junction, are used. All atomic positions along with the lattice constants are relaxed, till the forces on the atoms are less than 0.05 eV/\AA, with a total energy convergence criterion of $10^{-5}$ eV. 

For the PBE+U calculations we use the rotationally averaged formulation, as implemented in VASP, which applies a single parameter $U-J$ \cite{PhysRevB.57.1505}. We use a value $U-J=3$ eV, which is appropriate for $4d$ transition metals \cite{PhysRevB.82.195128,PhysRevB.86.165105}.
 
\section*{Results}
 
The macroscopic polarization is a topological invariant for 2D insulators with D$_{3h}$ symmetry\cite{PhysRevB.88.085110}. All semiconducting TMDCs that have the $H$ structure, belong to this class. Following the modern theory of polarization, it is straightforward to calculate the 2D polarization from first principles as an integral of the Berry phase over the Brillouin zone \cite{PhysRevB.47.1651}. For the $H$ structure, the direction of the polarization is normal to the zigzag direction of the hexagonal atomic pattern. Semiconducting TMDCs with the $T$ structure have zero polarization, as this structure is centro-symmetric.

Connecting two insulators with a different polarization at a junction results in a jump in the polarization, if the projection of the latter along the normal to the junction is nonzero. According to the theory of topological invariants, this must be accompanied by a closure of the band gap at the junction, or, in other words, the junction becomes metallic\cite{PhysRevB.88.161411,doi:10.1002/smll.201803040,C9CP01196J,PhysRevB.95.045302}. We call this a non-trivial junction. If one connects two insulators with the same polarization, or if the projection of the polarization along the surface normal is the same, then there is no change in the topological invariant. This leads to a trivial junction, which, in general, is insulating. 

In order to construct a non-trivial junction, we need to choose two TDMCs with different polarizations and connect them along a zigzag edge; an example is shown in Fig. \hyperref[figure1]{1a}. We also select TMDCs whose in-plane lattice constants are reasonably well matched, as too much strain at the junction would make the structure somewhat unrealistic from an experimental point of view.  In this paper, we will show a $H$/$H$ junction, which is of polar/polar type, and a $H$/$T$ junction, which is of polar/non-polar type, as examples of non-trivial heterojunctions in TMDCs. 

Going through a database of 2D TMDCs \cite{Mounet2018}, we identify a list of semiconducting $H$- and $T$-phase compounds. We then optimize their lattice constants and calculate their bulk polarization. In the $H$-phase, the entire group VI {\MX} (M = Mo,W; X = S,Se,Te) family of TMDCs has the same values for the polarization lattice, $\mathbf{P}=p_1\mathbf{a}_1+p_2\mathbf{a}_2$ with $(p_1,p_2)=(\alpha+n_1,\beta+n_2); n_{1,2}=0, \pm, 1,\pm 2 ... $, where $\mathbf{a}_{1,2}$ are the lattice vectors of the primitive 2D unit cell, and $(\alpha,\beta)$  has the value $(2/3,1/3)$. Choosing compounds for the junction from this group, for example, a heterojunction between MoS$_2$ and WS$_2$, means that the polarization is continuous across the interface, which results in a trivial, i.e., semiconducting interface. Several experiments and ab-initio calculations have indeed confirmed that this junction is semiconducting with a type-II (staggered) band alignment \cite{nmat4091,Zhang2018}. 

\subsection*{Polar/polar TMDC junctions}

\begin {figure}[htbp]
\begin{center}
\includegraphics[width=12.5cm]{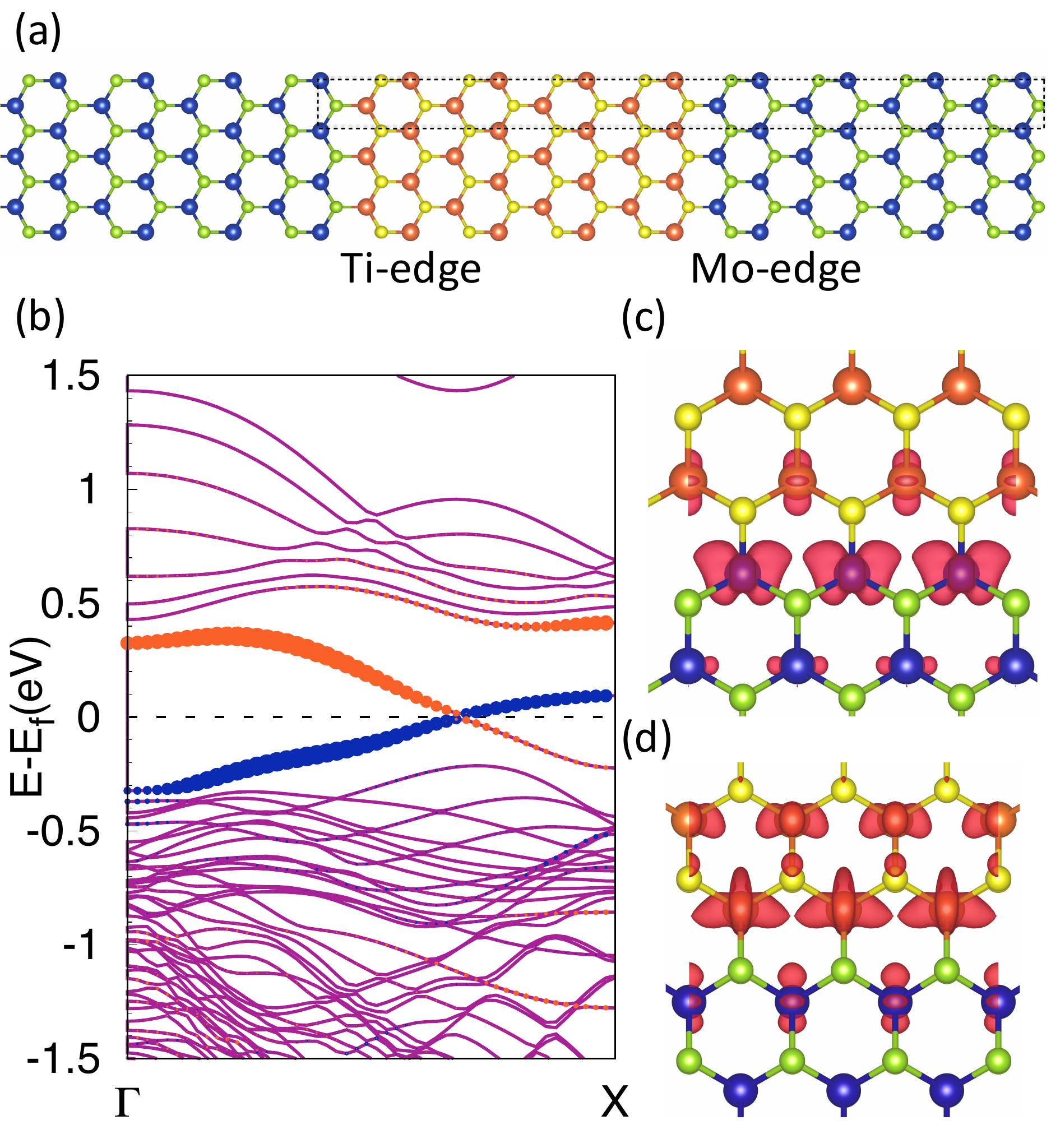}
\caption{ (a) Heterojunctions of $H$-{\MSe} and $H$-{\TiS}; the blue, green, orange and yellow spheres represent Mo, Se, Ti and S atoms respectively; the black dotted lines indicate the supercell. The two structurally different junctions are labeled Ti-edge and Mo-edge. (b) The band structure of the supercell; the states highlighted in blue and red are states belong to Mo and Ti edges, respectively, identified by projecting the wave function densities on the Mo and Ti atoms at the edges. (c) Wave function density at the Mo-edge at $E-E_F=-0.2$ eV (d). Wave function density at the Ti-edge at $E-E_F=+0.2$ eV.}
\end{center}
 \label{figure1}
\end{figure}

The group V TMDCs {\MX} (M = V,Nb,Ta; X = S,Se,Te) all are metallic, which is due to the transition metals having one less valence electron compared to group VI transition metals. Group IV TMDCs in the $H$ structure, such as TiX$_2$, are again semiconducting, with a calculated polarization value  $(\alpha,\beta)$=(1/3,2/3) that is different from that of group VI {\MX}. This suggests the possibility of creating a non-trivial, i.e., metallic, junction between TiX$_2$ and MX$_2$. The $T$ structure of TiX$_2$ is lower in energy than the $H$ structure, and is not suitable, as it is metallic, but we suggest that it may be possible to create the metastable $H$-structure by a suitable choice of growth conditions (as it is possible to create the metastable, metallic, $T$-structure of MoS$_2$ by suitable growth conditions).

To be specific, a $H$-{\TiS} monolayer has a calculated optimized lattice constant of 3.34 \AA, which is within 1\% of the optimized lattice constant of 3.31 \AA\ of {\MSe}. Both compounds are semiconductors, {\TiS} having a calculated (indirect) band gap of 0.7 eV, and {\MSe} having a (direct) band gap of 1.6 eV.  These two semiconductors form a type-II band alignment according to our DFT calculations, which, apart from the topological considerations discussed above, would result in a semiconducting junction. 

Fig. \hyperref[figure1]{1}(a) shows our supercell model for the {\TiS}/{\MSe} junction along the zigzag direction. There are two different junctions in the supercell, due to the periodic boundary conditions imposed in the calculations, We call call these the Ti-edge, respectively the Mo-edge, according to the transition metal atoms closest to the junction. Choosing a stoichiometric Se-S termination at the interfaces, as shown in the figure, we optimize the lattice constants again over the whole structure. After relaxation, the original bond lengths and angles are actually retained, and the structure has a uniform lattice constant close to 3.34 \AA.

Fig. \hyperref[figure1]{1}(b) shows the calculated non-spin-polarised bands of this structure, using a $1 \times$ periodicity in the direction of the Mo an Ti edges. There is a clear band gap, with two bands crossing the gap. The wave functions associated with these two bands are localized at the two junctions in the supercell, where one band can be assigned to the Ti edge, and the other to the Mo edge. The two edges clearly are metallic, the Ti edge band being $1/3$ occupied, and the Mo edge band $2/3$ occupied. 

The Mo edge state has dominant Mo $d_{xy}$ and $d_{z^{2}}$ character, see Fig. \hyperref[figure1]{1}(c), whereas the Ti edge state has foremost Ti $d_{xy}$ and $d_{x^{2}-y^{2}}$/$d_{z^{2}}$ character, with some participation of the other metal atom's $d$ states, see Fig. \hyperref[figure1]{1}(d). As the chalcogen atoms do not contribute appreciably to these edge states, this would imply that changing the chalcogen atoms at the edges would not affect the electronic structure. Indeed, on changing the chalcogen terminations at the junctions from a stoichiometric S-Se to a non-stoichiometric Se-Se or S-S, no change in the occupation or the dispersion of the edge states is observed. 

Both the existence of states localized at the junctions, with energies in the band gap, as well as the occupancy of those states can be deduced from the 2D polarization of the materials involved, and its topological character. Going from material 1 to material 2 across a junction, the polarization jumps from $\mathbf{P}_1$ to $\mathbf{P}_2$, which would result in a polarization line charge density at the junction $\lambda = (\mathbf{P}_2 - \mathbf{P}_1)\cdot \hat{\mathbf{n}}$, where $\hat{\mathbf{n}}$ is the in-plane unit vector normal to the junction. For a macroscopic junction this polarization charge has to be neutralized  by a compensating line charge density $\lambda_e=-\lambda$, such as to avoid an intrinsic electric field generating a polar catastrophy \cite{Nakagawa,PhysRevB.87.205423}. 

In the present case, the compensating charge can only be of electronic origin, and resides in states near the Fermi level, inside the band gap. In the language of topological invariants, going across a junction from an insulator with a certain value for the topological invariant to an insulator with a different value for that invariant, then at the junction the gap has to close, i..e, the junction is metallic, where the metallicity is carried by states localized at the junction.  

Going from {\MSe} to {\TiS}, this gives $\lambda_e = (2e)/(3a)$ at the Mo edge (where $a$ is the lattice constant along the edge, and $e$ is the elementary charge), which means that the corresponding edge state has to be $1/3$ occupied by holes (taking spin degeneracy into account). At the Ti edge, going from {\TiS} to {\MSe}, we have $\lambda_e = -(2e)/(3a)$, and an edge state that is $1/3$ occupied by electrons.

This interpretation seems similar to the three-dimensional case of junctions between insulating oxide perovskites, such as LaAlO$_3$/SrTiO$3$, where a 2DEG emerges at the interface between the two insulators \cite{oxide}. There, the interface metallicity has been attributed to the abrupt change of the valence charge of the cation at the interface, from La$^{3+}$ to Ti$^{4+}$, causing a charge transfer of 0.5 $e$ in order to avoid a polar catastrophy. The 2D heterojunction between AlN and SiC has been analyzed in similar terms, where the difference in formal charges between the cations and anions in the 2D III-V and the IV-IV materials drives the formation of a 1DEG at the interface, in order to avoid a similar polar catastrophy \cite{PhysRevB.88.161411}.

Our TMDC case is more subtle, because there is no ionic charge discontinuity at the interface. The formal charges on the cations and ions one both sides of the junctions are the same in their respective compounds, Ti$^{4+}$(S$^{2-}$)$_2$ and Mo$^{4+}$(Se$^{2-}$)$_2$. Moreover, the compounds have the same structure, and we also see no structural distortions at the junction. The difference in polarization between the two compounds is then of purely electronic origin, as is the metallicity of the junction.

Formally, the analysis of the metallic junction presented above only holds in the macroscopic limit, i.e., in the case where one has one junction between two semi-infinite 2D materials. Nevertheless, in our supercell model, which comprises strips of six TMDC units wide, we apparently already have reached this asymptotic limit. We find that if one uses strips of four or fewer units, one still observes edge states, see supporting information, but the edge states at the Mo edge and the Ti edge interact across the strip. This causes a gap to open up, similar to the bonding/anti-bonding interaction between atomic orbitals used to describe chemical bonding. 

We see no evidence for a residual electric field existing over the width of the ribbon. The electrostatic potential is approximately constant over the {\TiS} and the {\MSe} regions, and has a step at the interface, see supporting information. The electrons residing in the edge states have then fully compensated the polarization charge. This is in agreement with an earlier work that shows that an electric field is present only if there are no edge states to compensate for the differences in polarization \cite{PhysRevB.95.045302}. The band structure and the dispersion of the edge states suggest that the latter are not bulk states driven up or down by an electric field, but new states created in the gap \cite{PhysRevB.93.205444}.

\subsection*{Polar/non-polar TMDC junctions}

The $T$-phase structure of TMDCs has zero polarization because it has inversion symmetry. Most of the $T$-phase TMDCs are in fact metallic compounds, but {\HfS} and {\HfSe} are semiconductors. Constructing a junction between one of these and a semiconducting $H$--phase TMDC results in a polarization discontinuity and a metallic junction similar to that described in the previous section. The calculated lattice constant of $T$-{\HfS} is 3.62 \AA, which is close that of $H$-{\MTe}, 3.56 \AA, implying that a heterojunction between the two can be formed with minimal strain. The PBE band gaps of $T$-{\HfS} and $H$-{\MTe} are 1.3 eV and 1.1 eV, respectively, and the band alignment between the two is type II, which, apart from polarization considerations, would imply a semiconducting junction. 

\begin {figure}[htbp]
\begin{center}
\includegraphics[width=12.5cm]{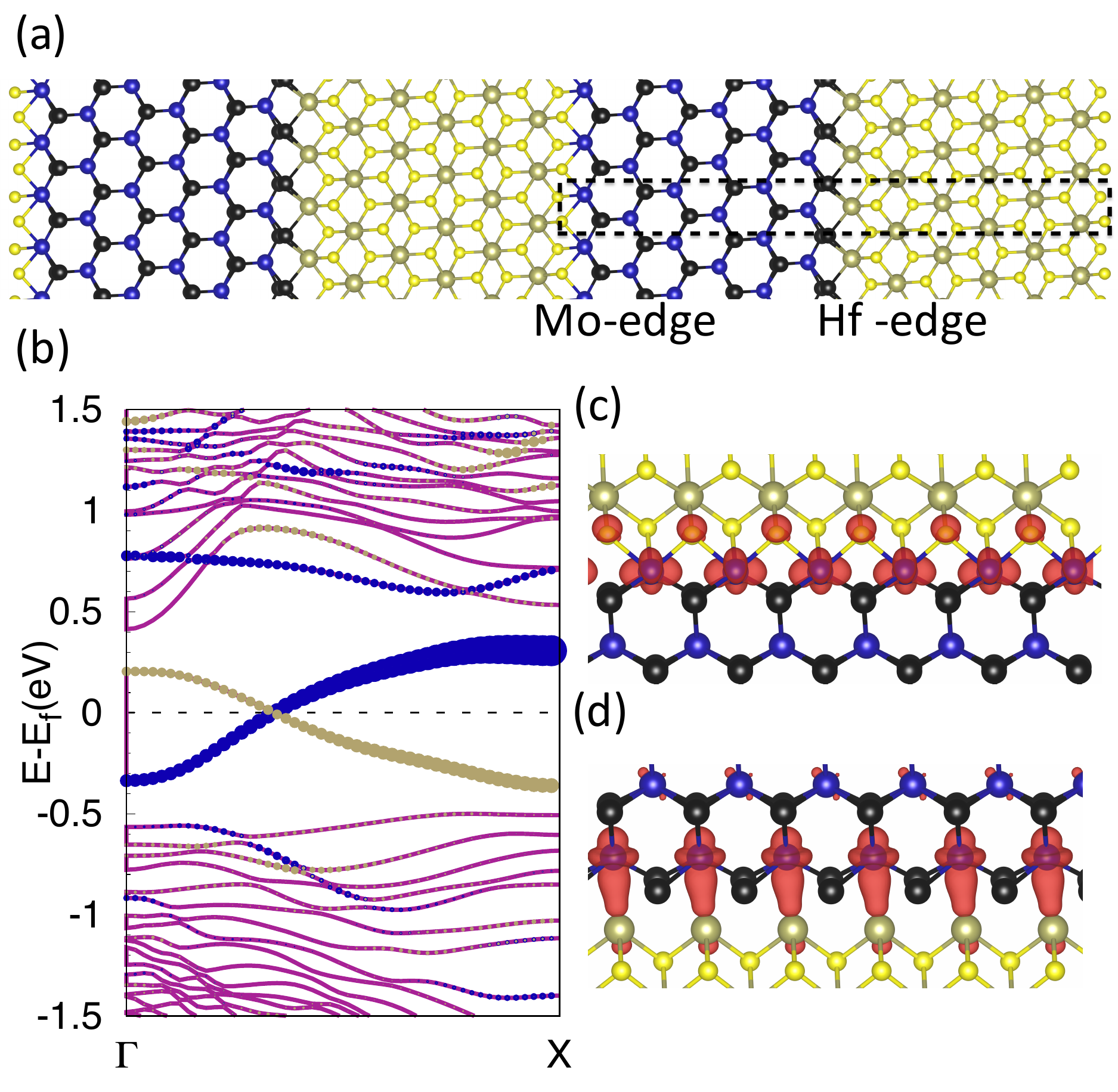}
\caption{ (a) Heterojunction of $H$-{\MTe} and $T$-{\HfS}; the blue, black, brown and yellow spheres represent Mo, Te, Hf and S atoms, respectively; the black dotted lines indicate the supercell. (b) The band structure of the heterojunction; the states highlighted in blue and brown are states belong to Mo and Hf edges, respectively, identified by projecting the wave function densities on the Mo and Ti atoms at the edges. (c) Wave function density at the Mo edge at $E-E_F=-0.3$ eV (d). Wave function density at the Hf edge at $E-E_F=-0.3$ eV.}
\label{figure2}
\end{center}
\end{figure}

$T$-$H$ junctions cannot be stitched perfectly, with all metal atoms at the junctions having sixfold coordination by chalcogen atoms. In experimental work one finds the $\beta$-junction structure \cite{C6RA26958C,Lin2014}, where the metal atom on the $T$ side of the junction (the Hf atom in this case) has sevenfold coordination, and the atom on the $H$ side (the Mo atom) has sixfold coordination. We adopt this structure at one interface, calling it the Hf edge in the following. At the other interface, called the Mo edge, we construct a similar structure,  giving the Hf atom a sixfold and the Mo atom a sevenfold coordination. Fig. \hyperref[figure2]{2}(a) shows the unit cell, where the two junctions are marked. More detailed images of the edge structures are given in the supporting information.

On optimizing, the lattice constant for the whole structure becomes 3.60 \AA. The bond lengths involving atoms close to the interface undergo changes, while the atoms far away from the interface in both the $H$- and $T$-phase remain at their respective bulk positions.

The coordination at the edges does not influence the bulk polarization, of course. It also does not affect the fractional character of the polarization charges at the edges, as adding or removing an atom simply adds or removes an integer number of electrons. Therefore, edge metallicity is robust against atomic defects at the junction. To test this we have also constructed a Mo-edge structure where the Hf atom has a fivefold and the Mo atom has a sixfold coordination, which gives a very similar electronic structure. Details can be found in the supporting information.

The electronic structure of $H$-{\MTe}/$T$-{\HfS} heterojunctions is shown in Fig. \hyperref[figure2]{2}(b). It clearly displays a band gap, with two bands, localized at the junctions, crossing in the gap at the fermi-level. The band from the Hf edge ($\beta$-junction) is $2/3$ occupied, and its character is a combination of Mo and Hf $d_{z^{2}}$ and $d_{x^{2}}$ orbitals,Fig. \hyperref[figure2]{2}(c). At the Mo edge, the band is $1/3$ occupied, and has predominant Mo $d_{xy}$ and $d_{z^{2}}$ character, see Fig. \hyperref[figure2]{2}(d). The existence of these localized states, and their occupancy follows from similar considerations as for the $H$-$H$ junctions discussed above.

\subsection*{1D electronic instabilities}

\begin{figure}[htbp]
\begin{center}
\includegraphics[width=12.5cm]{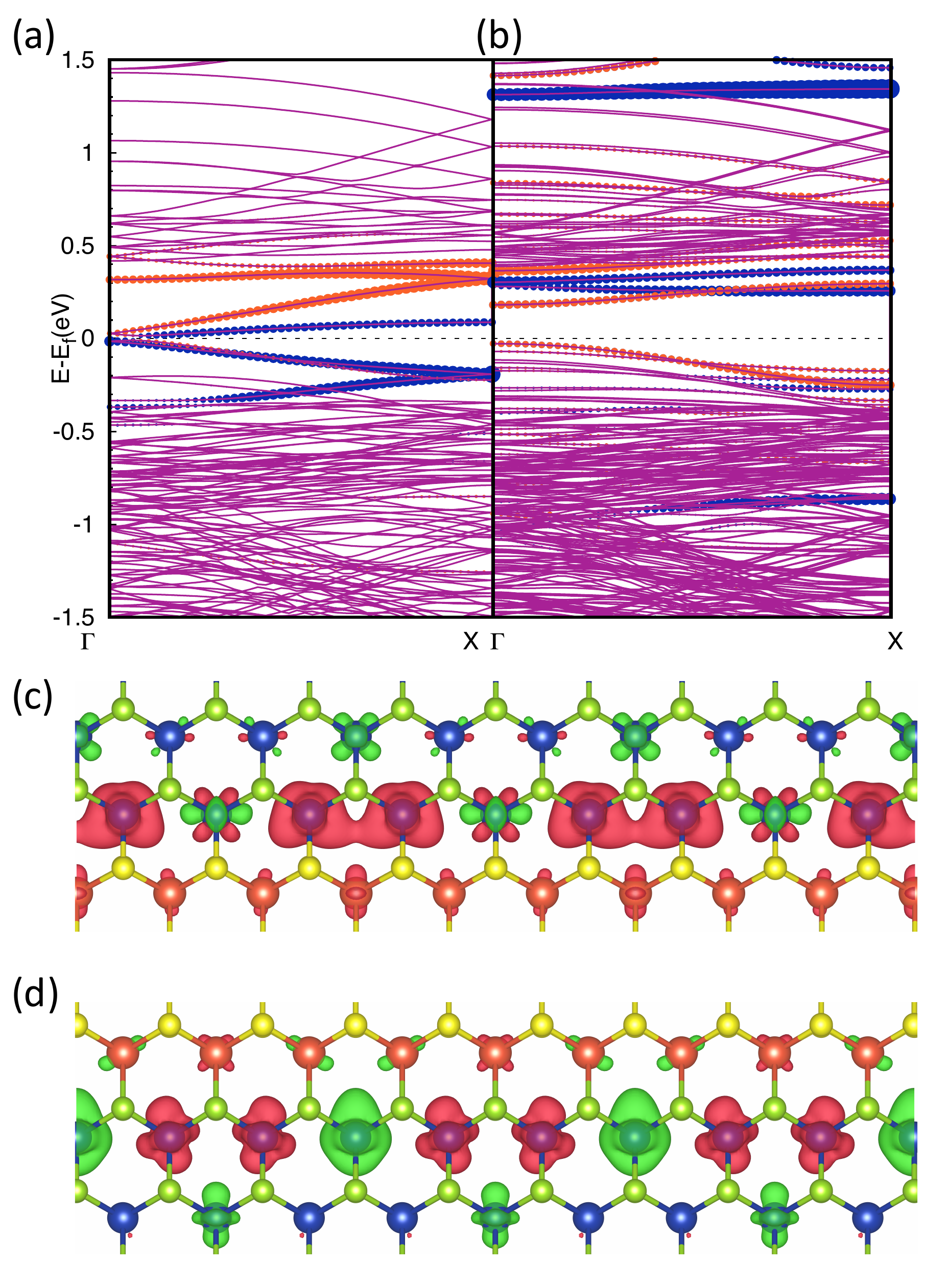}
\caption{The band structure of the $H$-{\MSe} and $H$-{\TiS} heterojunction calculated in the $3\times$ supercell; the states highlighted in blue and red are states belong to Mo and Ti edges, respectively. (a) 
Calculated with the PBE functional, and (b) with the PBE+U functional, both after structural relaxation. The latter gives a FM/AFM SDWs at the Mo/Ti edges, creating gaps of 0.45/0.22 eV, respectively. (c) Spin density wave at the Mo edge and (d) at the Ti edge; the red/green colors indicate spin up/down, where the SDWs result in the three Mo atoms at the junction becoming inequivalent \cite{vesta}.
}
\end{center}
 \label{figure3}
\end{figure}

The metallic states at TMDC junctions clearly have a 1D character. Electron correlation and electron-lattice interactions are particularly effective in 1D systems, and perturb the metallic character. In the present case, the 1D states have an occupancy of $1/3$ or $2/3$, which suggest the possibility of a CDW and/or a SDW that triples the period in the direction along the junction. This would be similar to the edges of a TMDC flake or a TMDC grain boundary that display a 3$\times$ periodicity \cite{Lucking,Barja:2016aa,Ma:2017aa,krishnamurthi2020,krishnamurthi2020edges}. 

We start from the $H$-{\TiS}/$H$-{\MSe} junction, triple our supercell along the direction of the junction, and reoptimize the structure, using the PBE functional. The structure in the $3\times$ cell does actually not change as compared to the $1\times$ structure, meaning that we do not observe a Peierls distortion. Fig. \hyperref[figure3]{3}(a) shows the $3\times$ band structure of the $3\times$ structure. Although the electronic structure close to the fermi level looks complicated, with multiple bands crossing, these bands are actually the same as the ones shown in  Fig. \hyperref[figure1]{1}(b), but folded because of the $3\times$ periodicity. 

The Mo-edge band, which is $2/3$ occupied in the $1\times$ cell, is then folded into three bands, with the two lower bands fully occupied, and the topmost one fully empty. The Ti-edge band, $1/3$ occupied in the $1\times$ cell, is folded similarly, with the lowest band fully occupied, and the two upper ones fully empty. This means also that no CDWs are observed in this calculation. Inclusion of spin polarization does not affect the interface or bulk bands, implying that also SDWs are absent.

Nevertheless, the metallic edge bands in Fig. \hyperref[figure3]{3}(a) display the prototypical 1D band structure that is susceptible to perturbations inducing a metal-insulator transition, such as a Peierls distortion, or a CDW/SDW, even though a DFT/PBE calculation does not find any of these. The 1D states have mostly Mo $d$ character, and although the on-site electron-electron Coulomb interaction in $4d$ transition metals is weaker than in $3d$ ones, it is not always negligible \cite{PhysRevB.82.195128,PhysRevB.86.165105}. In previous calculations on mirror twin boundaries of {\MSe} we have found that inclusion of on-site Coulomb and exchange interactions as in the PBE+U mean field approach, markedly changes the electronic structure of the 1D states.

Therefore, we repeat the calculations in the $3\times$ cell, using the PBE+U functional with $U-J=3$ eV for the Mo $4d$ electrons\cite{PhysRevB.82.195128,PhysRevB.86.165105}. In principle, one can also include such a parameter for the Ti $3d$ electrons, but this has little effect, as in {\TiS} these states are mostly empty. The top of the {\TiS} valence band has sulfur $p$ character, and the Ti $3d$ states only contribute significantly in the conduction band. We have also tested values of $U-J$ over the whole range 0-3 eV; details can be found in the supporting information.

The PBE+U functional indeed gives rise to the opening of a band gap in the band structure, as shown in Fig. \hyperref[figure3]{3}(b). It is caused by SDWs at both the Mo edge, as well as at the Ti edge junctions. In these SDWs the Mo atoms closest to the junctions carry a magnetic moment, whereas the Ti atoms at the junctions, and the atoms further away from the junctions, do not show any magnetic moments. Several (meta)stable magnetic configurations of Mo atoms at the two interfaces are found. The configuration with the lowest total energy has a ferromagnetic (FM) arrangement at the Mo edge, with magnetic moments of $1.0$, $1.0$ and $0.0$ $\mu_{B}$ on the three Mo atoms at the junction. An antiferromagnetic (AFM) arrangement is 60 meV/$3\times$ cell higher in energy. At the Ti edge we find the AFM arrangement to be lowest in energy, with magnetic moments on the three Mo atoms at the junction of $-1.0$, $+0.6$ and $+0.6$ $\mu_{B}$.

The SDWs are accompanied by CDWs and (slight) distortions of the structure at the junctions. The Mo atoms at the Mo edge adopt a 3$\times$ periodicity, with distances between the Mo atoms of 3.27 \AA\ and 3.41 \AA, whereas the distances between the Ti atoms at the Mo edge do not change much. At the Ti edge, the distances between Ti atoms are 3.29 \AA\ and 3.33 \AA, and there are no significant changes in the bond distances between the Mo atoms.


The SDW/CDWs lead to a metal-insulator transition, resulting in a band gap in both the spin configurations. The band gap at the Mo edge is  $0.45$ eV whereas at the Ti edge, it is $0.22$ eV, this lowers the total energy by 330 meV/3$\times$ cell compared to the undistorted structure. The exact value of $U-J$ is not too critical for the emergence of a SDW/CDW and a gap. The SDW/CDW persists for values of $U-J$ in the range 2-3eV, although the magnetic moments and the band gap decrease on decreasing $U-J$, see supporting info. The correlation-driven SDW/CDWs seem to be a unique feature of 2D TMDC junctions, due to the presence of $d$ electrons on the transition metals. In the AlN-SiC case, where $d$ electrons are absent, spin-polarised DFT calculations do not give SDW/CDWs, and the junctions remain metallic  \cite{PhysRevB.88.161411}. 

\begin {figure}[htbp]
\begin{center}
\includegraphics[width=12.5cm]{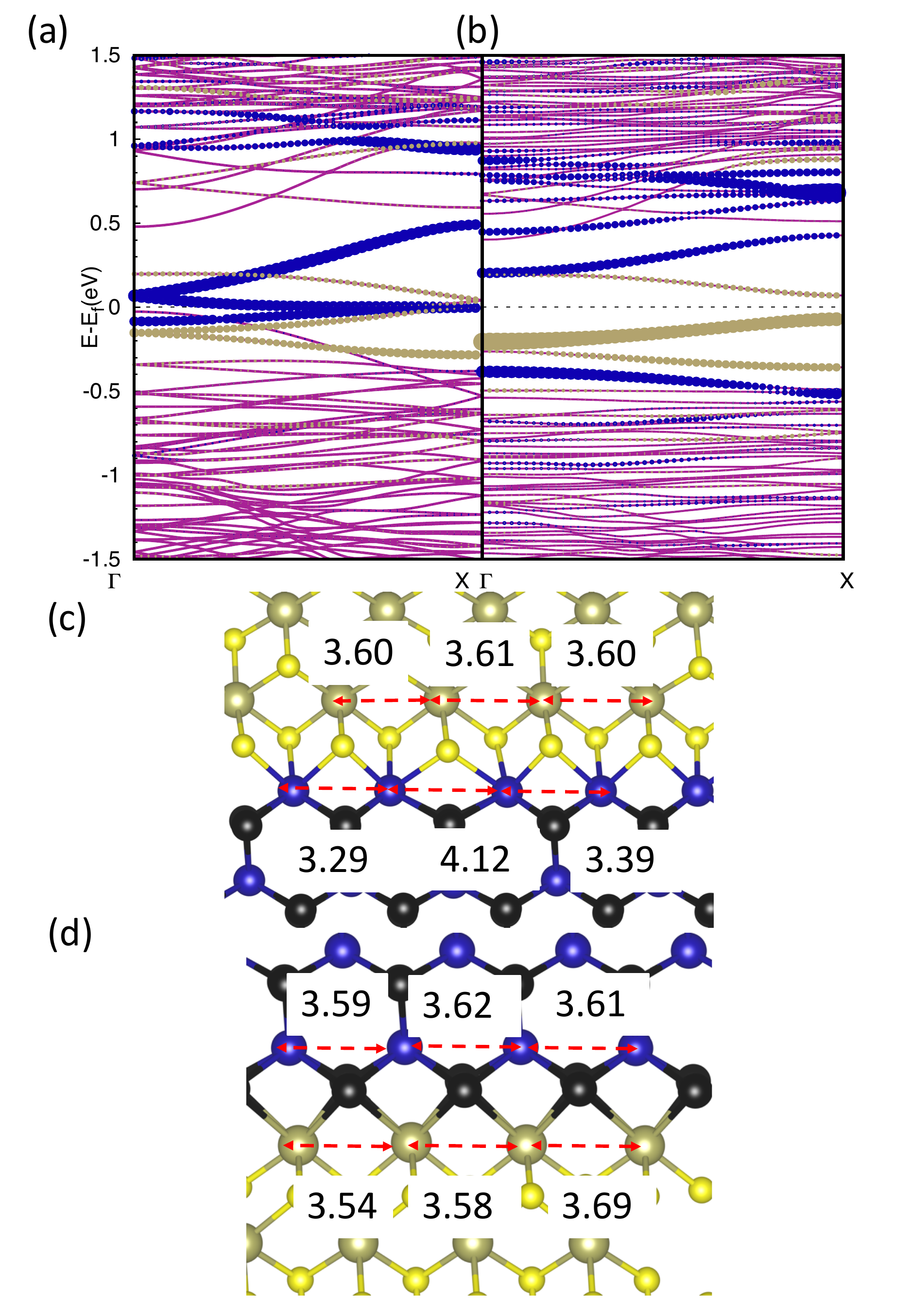}
\caption{ The band structure of the $H$-{\MTe} and $T$-{\HfS} heterojunction calculated in the $3\times$ cell; the states highlighted in blue and brown are states belong to Mo and Hf edges, respectively. (a) undistorted $1\times$ structure, (b) Peierls distorted structure; band gaps at the Hf and Mo edges are 0.15 eV and 0.6 eV, respectively. Peierls distorted structures at (a) Mo edge, and (d) Hf edge; the numbers give the distances in \AA\ between metal atoms at the interfaces.}
\end{center}
 \label{figure4}
\end{figure}

In the case of the $H$-{\MTe}/$T$-{\HfS} junction, simply optimizing the 3$\times$ structure with the PBE functional gives a distinct CDW with a clear $3\times$ modulation of the structure at both the junctions. The modulation is largest at the Mo edge, with distances between the Mo atoms at the edge becoming 3.39 \AA\ and 4.11 \AA. At the Hf edge the modulation is somewhat smaller, with distances between the Hf atoms at the edge becoming 3.54 \AA\ and 3.69 \AA. These Peierls distortions open up a gap of 0.6 eV at the Mo edge and 0.15 eV at the Hf edge, and lower the energy by 270 meV/3$\times$ cell. The band structures of the undistorted and the distorted structures are shown in Fig. \hyperref[figure4]{4}

Adding spin polarization has no effect, but upon adding on-site interactions, with $U-J=1.5$ eV, magnetic moments of $-0.48$, $0.54$ and $-0.30$ $\mu_{B}$ develop on the Mo atoms at the Mo edge, whereas the Hf edge remains unpolarized. The band gaps change only slightly upon adding U; at the Hf edge it is 0.17 eV and at the Mo edge it is 0.42 eV, see supporting information for the corresponding band structures.

\section*{Summary and conclusions}

In summary, we have shown by means of first-principles DFT calculations that lateral heterojunctions of 2D semiconducting TMDCs can have 1D metallicity in a 1$\times$ unit cell. The metallicity arises out of a discontinuous topological invariant at the junction - the in-plane electric polarization, which is nonzero for TMDCs in the $H$-structure. Such heterojunctions can be made from two semiconducting TMDCs, both in the $H$ phase, with different polarization, and matching lattice constants. Alternatively, junctions can be made between a TMDC in the $H$-structure with nonzero polarization, and one in the $T$-structure with zero polarization, again with matching lattice constants. Changing the chalcogen atoms or creating defects in at the junctions will not affect the bulk polarization and hence metallicity is preserved. 

The 1D metallic states are however susceptible to the instabilities of 1D metals originating from electron-electron and electron-lattice interactions. Using DFT+U calculations, we show that spin- and/or charge-density waves can create a band gap at the junction, which is, however, much smaller than the band gaps of the two TMDC semiconductors. The details of these density waves depend upon the detailed structure of the junction. We propose that 2D TMDC heterojunctions are ideal systems for studying 1D physics.

\section*{Acknowledgements}
This work was financially supported by the \enquote{Nederlandse Organisatie voor Wetenschappelijk Onderzoek} (NWO) through the research programme of the former \enquote {Stichting voor Fundamenteel Onderzoek der Materie} (NWO-I, formerly FOM) and through the use of supercomputer facilities of NWO \enquote{Exacte Wetenschappen}(Physical Sciences). We acknowledge the funding from the Shell- NWO/FOM Computational Sciences for Energy Research program (Project No. 15CSER025)

\section*{References}
\bibliography{/Users/sridevi/Desktop/Lateral-Junction-paper/2D_LHS/LHS_2d.bib}

\end{document}